# Non-conventional phase attractors and repellers in weakly coupled autogenerators with hard excitation

Margarita Kovaleva[1]; Leonid Manevitch[1]; Valery Pilipchuk[2]

[1]N.N. Semenov Institute of Chemical Physics
Kosygin st., 4 Moscow,Russia 119991

e-mail address: makovaleva@chph.ras.ru,

manevitchleonid3@gmail.com

[2]Wayne State University,
1200 Holden Street, MI 48202, Detroit, USA

e-mail address: pilipchuk@wayne.edu

**Abstract** In our earlier studies, we found the effect of non-conventional synchronization, which is a specific type of nonlinear stable beating in the system of two weakly coupled autogenerators with hard excitation given by generalized van der Pol-Duffing characteristics. The corresponding synchronized dynamics are due to a new type of attractor in a reduced phase space of the system. In the present work, we show that, as the strength of nonlinear stiffness and dissipation are changing, the phase portrait undergoes a complicated evolution leading to a quite unexpected appearance of difficult to detect 'repellers' separating a stable limit cycle and equilibrium points in the phase plane. In terms of the original coordinates, the limit cycle associates with nonlinear beatings while the stationary points correspond to the stationary synchronous dynamics similar to the so-called nonlinear local modes.

1. **Introduction**

The synchronization of coupled oscillators with similar characteristics is a fundamental phenomenon, which can be developed in very different natural systems regardless their physical contents [1-7]. Since its first observation by Huygens in the system of two pendulum clocks hanging on the wall, the synchronization was always associated with almost constant amplitudes and phase shifts. Typically such oscillations are synchronized either in-phase or anti-phase similarly to the so-called Nonlinear Normal Modes (NNMs) motions of conservative systems [8]. In particular, Huygens observed the anti-phase synchronization [1], when all the system behaves as a 'collective' single degree-of-freedom oscillator. The similarity with NNMs provides the tool for system reduction and thus explains the possible reason why the conventional synchronization was mainly in focus of previous studies. However, recently we found a new type of synchronization accompanied by very intensive energy exchange between the oscillators as described in [9, 10] and [13]. This type of synchronization represents a stable nonlinear beating between the generators, when the phase shift remains quite close to a constant almost all of the time, except for relatively short periodic jumps between in-phase and anti-phase states. In this case, due to the fact that amplitudes are strongly modulated, the well-known phase approximation [2, 6] does not provide a complete description of main dynamic specifics. In contrast to the conventional NNMs synchronization, the distribution of excitation between the system's oscillators is non-stationary. Moreover, as mentioned above, it is rather close to the extreme case with a maximal exchange of the excitation. As a result, the NNMs reduction is not applicable any more, whereas an alternative limit case emerges according to the idea of Limiting Phase Trajectories – LPTs [11, 12]. In conservative cases, the LPTs are boundaries of phase cells along which the entire energy cyclically swings from one oscillator to another usually with non-smooth temporal shapes. Note that such shapes admit analytical approximations using the non-smooth basis given by dynamic states of impact oscillators with stiff barriers [16]. The LPT type 'nonstationary synchronization' appears as an attractor in the non-conservative system of quasilinear active oscillators near the boundary of phase cells. We have also observed numerically similar stable oscillations in a system with more degrees of freedom, when the weak coupling between two active oscillators is realized via linear oscillator [14]. Furthermore, it was shown that similar mechanism determines the superradiant transition in the quantum two-level system in the presence of electromagnetic field whose role is similar to that of the linear oscillator in the classical case. This similarity gives an unexpected and useful example of quantum-classical analogy [15].

In the present work, we investigate the topological evolution of phase flows surrounded by such attractors as the strength of conservative nonlinearity increases. While the stable limit cycle represents the LPT-type synchronization, the stationary points represent the conventional type of modal synchronization with fixed phase shifts.

## 2. Model of coupled active oscillators and its reduction

We consider the model of two generalized Van der Pole–Duffing nonlinear oscillators:

$$\frac{d^2u_1}{dt^2} + u_1 + 8\alpha\varepsilon u_1^3 + 2\beta\varepsilon(u_1 - u_2) + 2\varepsilon(\gamma - 4bu_1^2 + 8du_1^4)\frac{du_1}{dt} = 0;$$
$$\frac{d^2u_2}{dt^2} + u_2 + 8\alpha\varepsilon u_2^3 + 2\beta\varepsilon(u_2 - u_1) + 2\varepsilon(\gamma - 4bu_2^2 + 8du_2^4)\frac{du_2}{dt} = 0.$$ (1)

Following the procedure applied to the system in our previous work [9] we introduce complex variables $\psi_j = v_j + iu_j$, $v_j = du_j/dt$ ($j = 1,2$) or $v_j = (\psi_j + \psi_j^*)/2$ and $u_j = -i(\psi_j - \psi_j^*)/2$, where the asterisk indicates complex conjugate, and proceed further to the two-variable expansions technique by introducing two different temporal scales as $\psi_j = [\varphi_{j,0}(\tau_0, \tau_1) + \varepsilon\varphi_{j,1}(\tau_0, \tau_1) + O(\varepsilon^2)]e^{i\tau_0}$; $j = 1,2$, where $\tau_0 = t$ and $\tau_1 = \varepsilon t$ are fast and slow time variables, respectively. Then, applying the substitution $\varphi_{j,0} = f_j e^{i\beta\tau_1}$ gives the leading-order asymptotic approximation in the complex form:

$$\frac{df_1}{d\tau_1} - 3i\alpha |f_1|^2 f_1 + (\gamma - b|f_1|^2 + d|f_1|^4) f_1 + i\beta f_2 = 0,$$
$$\frac{df_2}{d\tau_1} - 3i\alpha |f_2|^2 f_2 + (\gamma - b|f_2|^2 + d|f_2|^4) f_2 + i\beta f_1 = 0. \qquad (2)$$

As shown in [9-10], system (2) possesses the integral of motion $N = |f_1|^2 + |f_2|^2 = 2b/(3d)$ if the system parameters satisfy the relationship $b^2 = 9\gamma d/2$. In order to describe the dynamics along the integral manifold, we further use the new angle coordinate $\theta(t)$ according to the relationships $f_1 = \sqrt{N}\sin\theta e^{i\delta_1}$ and $f_2 = \sqrt{N}\cos\theta e^{i\delta_2}$. This brings system (2) to the final form

$$\frac{d\theta}{d\tau_2} = \frac{1}{2}(\sin\Delta - \lambda\sin 4\theta),$$
$$\sin 2\theta \frac{d\Delta}{d\tau_2} = \cos 2\theta \cos\Delta + 2k\sin 4\theta \qquad (3)$$

where $\Delta = \delta_2 - \delta_1$, $\tau_2 = 2\beta\tau_1$ is a new time-scale, and parameters $k = 3\alpha N/(8\beta)$ and $\lambda = N^2 d/(8\beta)$ characterize the conservative part of nonlinearity and dissipation, respectively.

### 3. Phase plane analysis

For comparison reason, the left column of Fig.1 illustrates the evolution of the phase plane in the conservative case, λ=0, whereas the right column of the phase plane shows the related phase portraits in the dissipative case, λ≠0, under the same nonlinearity levels.

*Conservative system: weakly coupled oscillators with quasilinear nonlinearity.*

Let us discuss first the phase planes in the conservative case of weakly coupled quasilinear oscillators. The phase plane consists of two periodically repeated cells, however Fig. 1 shows only the 'elementary' fragment of the phase plane including two qualitatively different cells. At low nonlinearity levels, there are two stationary points, (0, π/4) and (π, π/4), at the centres of both cells, representing the in-phase and anti-phase NNMs, respectively. The cell boundaries separating the cells of in-phase and anti-phase modes represent specific trajectories – Limiting Phase Trajectories (LPTs). As noticed in Introduction, the LPT is associated with the complete energy transfer between the oscillators, and thus can play the role of a 'nonstationary alternative' to stationary NNM dynamic regimes. There are two subsequent topological transitions that can be observed in the phase plane. The first one occurs when the nonlinearity parameter $k$ exceeds the value ¼. At this nonlinearity level, the anti-phase NNM becomes unstable and two new stationary points are born. From the physical standpoint, such a bifurcation points to the onset of energy localization on individual oscillators, which is the so-called nonlinear local modes (NLMs). This happens due to the fact that high nonlinearity levels make the oscillators effectively stiffer or, in other words, the coupling between them weaker. Note that transition to the non-conservative case makes the notion of energy somewhat vague. Nevertheless, the energy term of the corresponding conservative oscillator still can be viewed as a Lyapunov's function describing the excitation level in the non-conservative case.

*Dissipative case: weakly coupled autogenerators*

The column on the right of Fig. 1 presents phase portraits of system (3) under a gradually increasing nonlinearity level while the dissipation parameter is fixed, λ>0. As seen from Fig.1, all the topological transformations happen inside the cell of anti-phase mode. Due to the presence of non-conservative term, the NNM stationary point becomes unstable focus while the nonlinearity level is low enough. In this case, a limit cycle, which is very close to the cell boundary, becomes the only attractor of the system representing the non-stationary LPT-type synchronization. *Note that the term 'limit cycle' relates to the reduced system (3) describing the slow-time modulation dynamics of the original system (1).* During this type of synchronization, the "energy" is periodically transferred from one oscillator to another while the phase shift between the oscillators Δ remains almost constant all the time except for very narrow time intervals. During such interval the phase shift Δ jumps along the horizontal pieces of limit cycle while the "energy" distribution, described by the angle $\theta(t)$ is almost fixed. As mentioned in Introduction such temporal behaviours admit approximations with the non-smooth periodic basis [16]. Note that the limit cycle can be attractor provided that the nonlinear dissipation parameter satisfies the relationship $\lambda < \left(1+\sqrt{1-4k^2}\right)/2$ [9, 13]. Now we focus on the evolution of the phase plane assuming the nonlinearity parameter can vary within a relatively large range. We showed that the evolution of the phase flow inside the cell of anti-phase mode develops in several stages as follows; see the right column of Fig. 1. In particular, increasing the nonlinearity parameter *k* above the anti-phase NNM's stability threshold still keeps the limit cycle corresponding to nonconventional synchronization stable, although the two new NNMs are seen bifurcated from the antiphase mode; Fig.1(d). Since the newborn normal modes are stable focuses, two unstable limiting cycles (repellers) encircling each of the newborn modes develop; see the red curves in Fig. 1(f). Since the both repellers occur from the separatrix loops, the dynamics in their vicinities are *extremely slow* compared to those close to the larger (stable) limit cycle.

If we further increase the nonlinearity parameter the repellers collide with each other to form a new (unstable) limit cycle that separates two areas of attraction; see the red curve in Fig.1(h). Outside the repeller loop, the non-conventional synchronization is possible, while inside the repeller synchronization on one of the two localized NNMs is developed. In Fig.2, we illustrate the behavior of system (1) using the set of parameters, corresponding to Fig 1(h). In particular, Fig.2(a) shows the dynamics under the initial conditions inside the repeller. It is seen that most of the energy/excitation becomes localized on the second oscillator. When the initial conditions are taken outside the repeller, the system becomes attracted to the beat-wise dynamics with the intensive energy exchange; see Fig. 2(b). Therefore, the repellers separate domains of attraction of different types of synchronization.

Eventually the stable and unstable limiting cycles collide and the newborn focuses become the only attractors of the system; see Fig.1(j). These attractors correspond to a synchronization with predominantly one of the two oscillators excited.

## 4. New-born asymmetrical modes and their stability

To predict the parameter range for the existence of repellers we study the stability of the stationary points in the neighbourhood of anti-phase mode. Such stationary points can be interpreted as additional nonlinear normal modes with partial localization of excitation on one of the two oscillators. Note that the coordinates of both points admit exact analytical solutions

$$\theta_1 = \frac{1}{2}\arcsin\left(\sqrt{\frac{1}{2}\left(1 + \frac{4k^2}{\lambda^2} - \sqrt{\left(1 + \frac{4k^2}{\lambda^2}\right)^2 - \frac{1}{\lambda^2}}\right)}\right),$$

$$\Delta_1 = \pi - \arccos\left(4k\sqrt{\frac{1}{2}\left(1 + \frac{4k^2}{\lambda^2} - \sqrt{\left(1 + \frac{4k^2}{\lambda^2}\right)^2 - \frac{1}{\lambda^2}}\right)}\right)$$

(4)

and

$$\theta_2 = \frac{\pi}{2} - \frac{1}{2}\arcsin\left(\sqrt{\frac{1}{2}\left(1 + \frac{4k^2}{\lambda^2} - \sqrt{\left(1 + \frac{4k^2}{\lambda^2}\right)^2 - \frac{1}{\lambda^2}}\right)}\right),$$

$$\Delta_2 = \pi + \arccos\left(4k\sqrt{\frac{1}{2}\left(1 + \frac{4k^2}{\lambda^2} - \sqrt{\left(1 + \frac{4k^2}{\lambda^2}\right)^2 - \frac{1}{\lambda^2}}\right)}\right).$$

(5)

The presence of solutions (4) and (5) enables us to linearize system (3) near the stationary points and thus conduct their local stability analysis. Note that eigen values of the corresponding Jacobian matrixes appear to be the same for both points due to the symmetry. In other words, points (4) and (5) possess the same stability properties as soon as nonlinearity and dissipation parameters of system (3) remain fixed. The result of such analysis is summarized in Fig. 3, where the curve represents a manifold of parameters $\{\lambda,k\}$ that separates the areas of positive and negative Lyapunov's exponents in such a way that they are positive below the curve. Therefore, points (4) and (5) are stable above while unstable below the curve in Fig. 3. Direct numerical integrations of equations (3) on both sides of the boundary are illustrated in Fig. 4 to confirm our conclusion.

*Remark*: In the vicinity of the antiphase mode the system (3) can be represented in the form:

$$\frac{dx}{d\tau_2} = 2(1-4k)y + \frac{8}{3}(1+2k)y^3 - x^2 y$$

$$\frac{dy}{d\tau_2} = -\frac{1}{2}x + 2\lambda y + \frac{1}{12}x^3 - \frac{16}{3}\lambda y^3$$

(6)

where $\Delta = \pi + x$, and $\theta = \pi/4 + y$. As is seen from the Figure 6 the system (6) represents evolution of all the (stable and unstable) limit cycles of the system (3). When the unstable limit cycle are born, the stationary points inside each of them are stable focuses.

## 5. The effect of soft and stiff nonlinearities

In the previous sections we studied the case of the "hard" nonlinearity. If the sign of the parameter $α$ is changed to negative, the effect of the conservative nonlinearity is changed quite drastically from the standpoint of its effect on the evolution of the phase planes. Let us note that all the analysis provided for the case of 'stiff' nonlinearity can be completed for the 'soft' one as well. The new asymptotic system will coincide with the system (3). Only the sign of the parameter $k$ must be changed. Therefore, the phase plane analysis remains applicable to the case of soft nonlinearity leading however to quite different outcome. For instance, the anti-phase NNM will remain stable in all the range of parameters considered, while the in-phase NNM becomes unstable if the threshold value $k=¼$ is exceeded. In the non-conservative case, the stable limit cycle with the intensive energy exchange occurs around the in-phase mode. All the phase plane analysis can be conducted similarly to the 'stiff' case. The only difference is that the in-phase and anti-phase mode cells will replace each other o the phase plane; see Fig.7 for illustration.

## 6. Conclusions

In the present work, we revealed all possible types of synchronization in quasilinear system of two weakly coupled active oscillators with hard excitation. It was shown that the increase of conservative nonlinearity is accompanied by quite a complicated evolution of possible regimes of synchronization. We described all the stages of evolution in terms of phase portraits of the asymptotically reduced model. This may provide an adequate tool for controlling the synchronized dynamics of weakly coupled generators through appropriate choice for the system parameters and initial conditions.

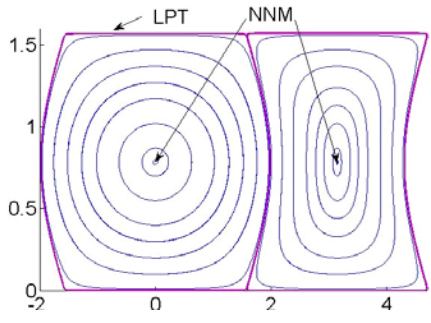

(a)

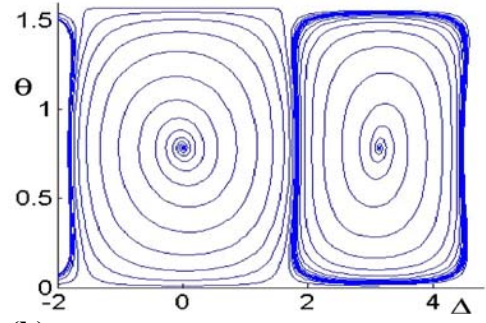

(b)

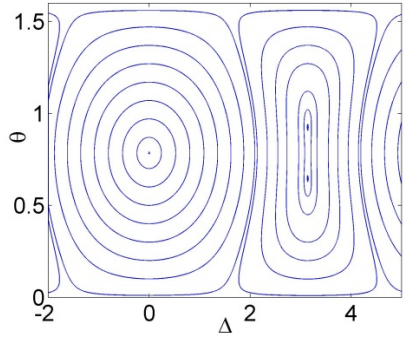

(c)

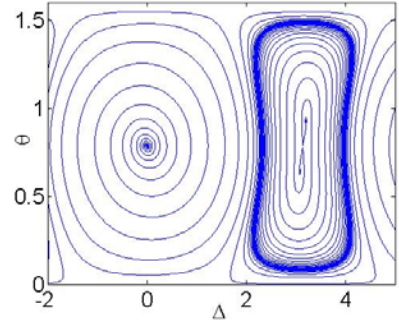

(d)

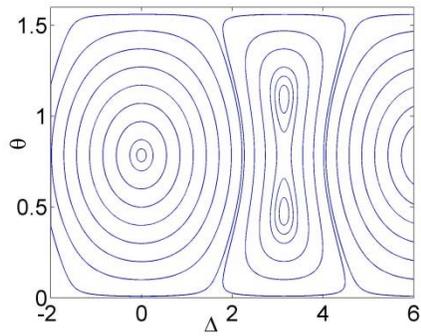

(e)

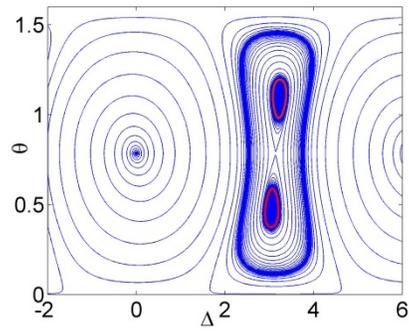

(f)

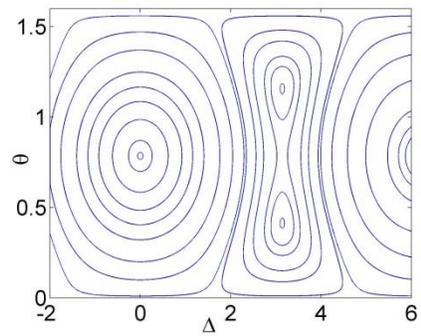

(g)

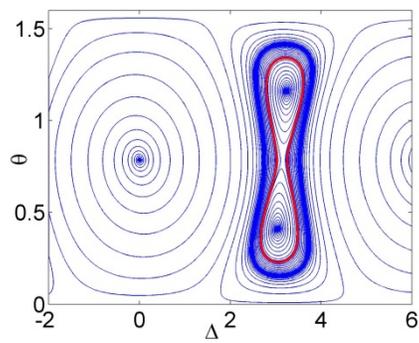

(h)

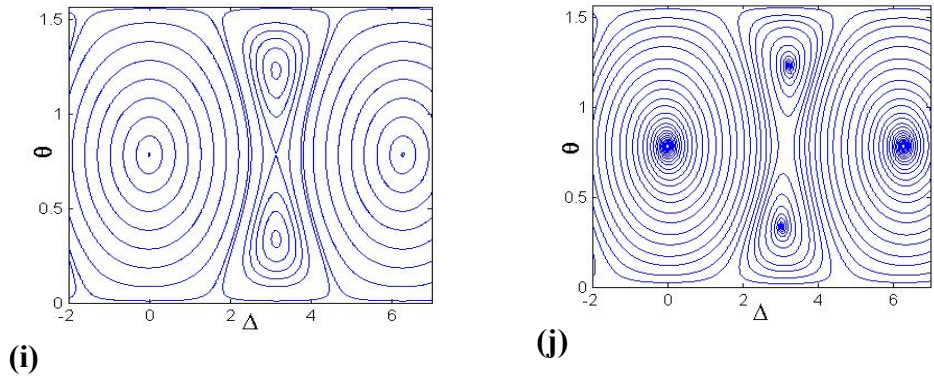

Figure 1. Evolution of the phase plane of system (3) in the 'weakly nonlinear' case, due to a gradual nonlinearity increase, where the variables $\theta$ and $\Delta$ characterize the relation between the two amplitudes and the phase shift, respectively: (a) nonlinearity $k = 0.2$, the dissipation $\lambda = 0$ - a conservative system; (b) $k = 0.2$ and $\lambda = 0.1$ - weak dissipation, the stationary points are unstable focuses, the limit cycle corresponding to LPT becomes attractor; (c) $k = 0.26$ and $\lambda = 0$ - the out-of-phase NNM becomes unstable, two new stationary points are born; (d) $k = 0.26$ and $\lambda = 0.1$ - the limit cycle is still stable; (e) $k = 0.31$ and $\lambda = 0$; (f) $k = 0.31$ and $\lambda = 0.1$ - further increase of the nonlinearity does not affect the topology of phase portrait, however, the limit cycles encircling the newborn stationary points can be observed; (g) $k = 0.34$ and $\lambda = 0$; (h) $k = 0.34$ and $\lambda = 0.1$ - two unstable limit cycles collide by forming one unstable limit cycle, attractors of the system remain same; (i) $k = 0.4$ and $\lambda = 0$; (j) $k = 0.4$ and $\lambda = 0.1$ - stable and unstable limit cycles collided, the two stable focuses become the only attractors of the system.

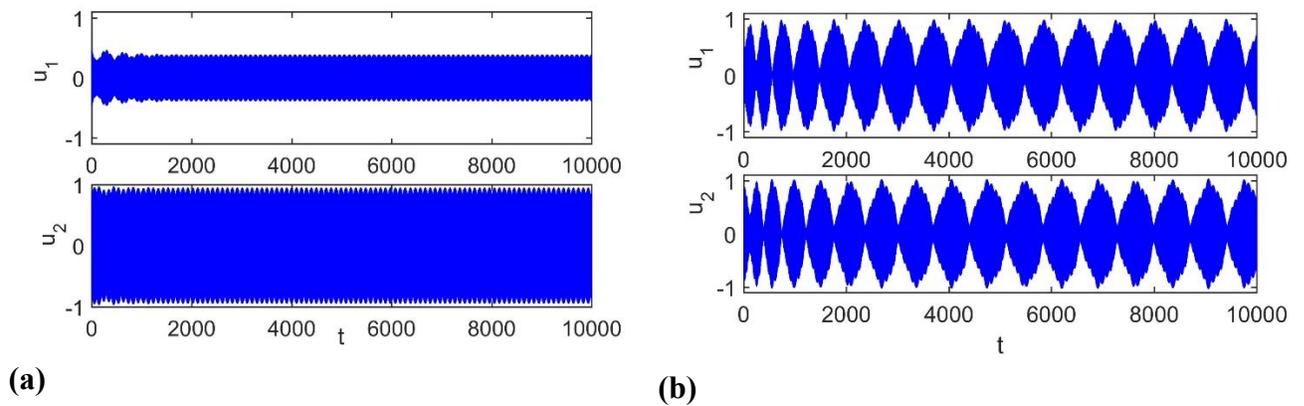

Figure 2. Numerical solutions of equations (1): a) inside of the repeller: $u_1(0) = 0.4794; v_1(0) = 0$; $u_2(0) = -0.8776; v_2(0) = 0$; b) outside of the repeller: $u_1(0) = 0.4794; v_1(0) = 0$; $u_2(0) = 0.8776; v_2(0) = 0$; the parameter values are: $\varepsilon = 0.01$; $\alpha = 0.9071$; $\beta = 1.0$; b=2.4; d=0.8; k=0.34; $\lambda$=0.1

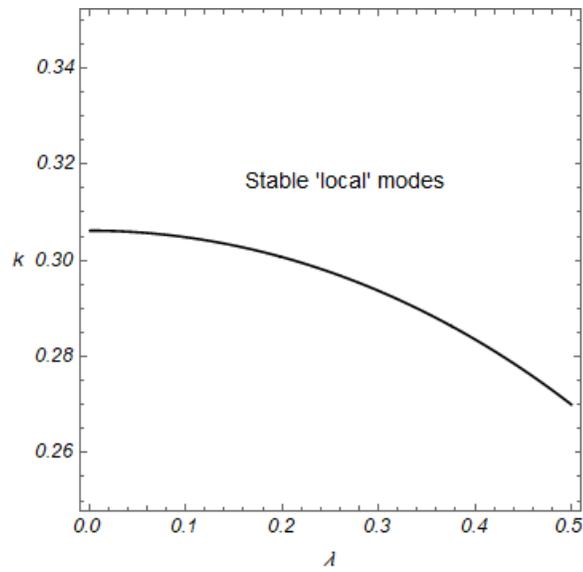

Figure 3. The boundary of stable quasi local modes in the nonlinearity ($k$) – dissipation ($\lambda$) plane.

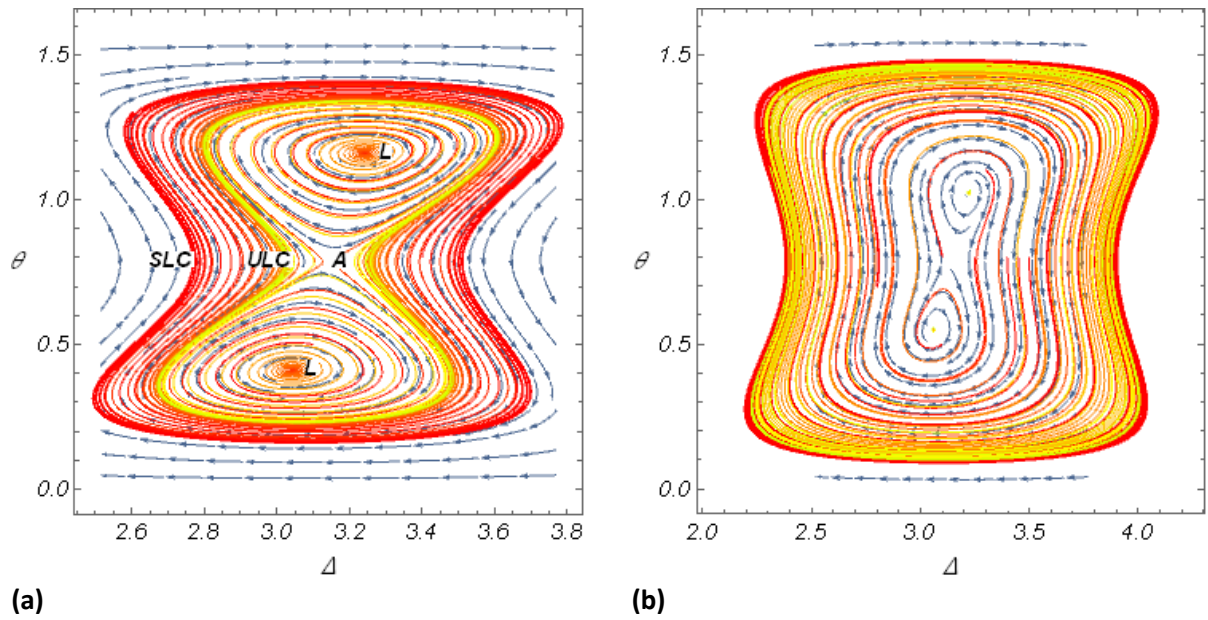

(a)                                                                 (b)

Figure 4. Phase portraits of system (3) above (a) and below (b) the boundary, which is shown in Fig.3: (a) $k = 0.34$, $\lambda=0.1$; (b) $k = 0.28$, $\lambda=0.1$; in particular, the fragment (a) shows the saddle point $A$, corresponding to the unstable antiphase mode, two stable spiral points $L$, corresponding to stable 'quasi local' modes, the unstable $ULC$ and stable $SLC$ limit cycles.

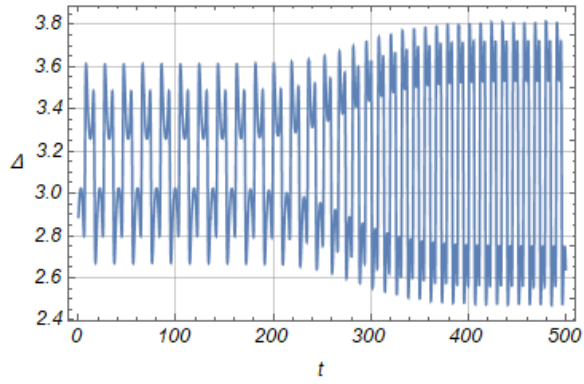
(a)

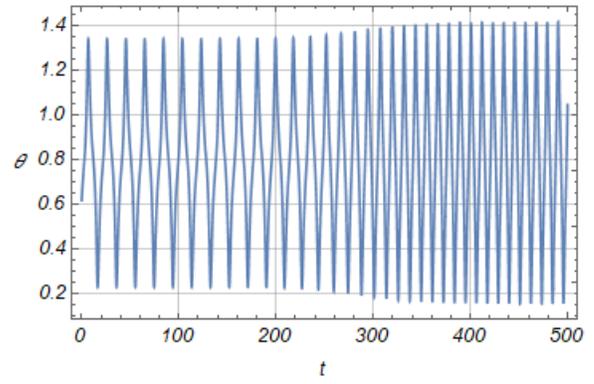
(b)

Figure 5. Time history of the transition from the unstable *ULC* to stable *SLC* limit cycles; see the phase portrait in Fig. 4a.

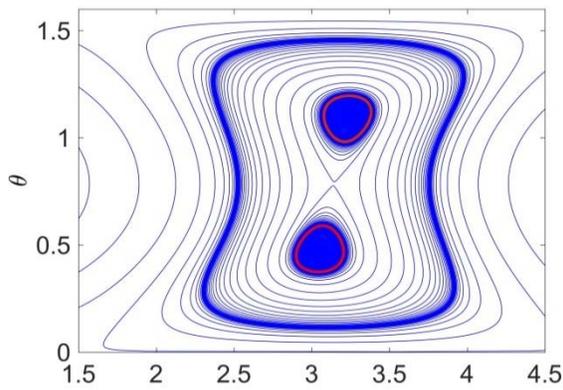
(a)

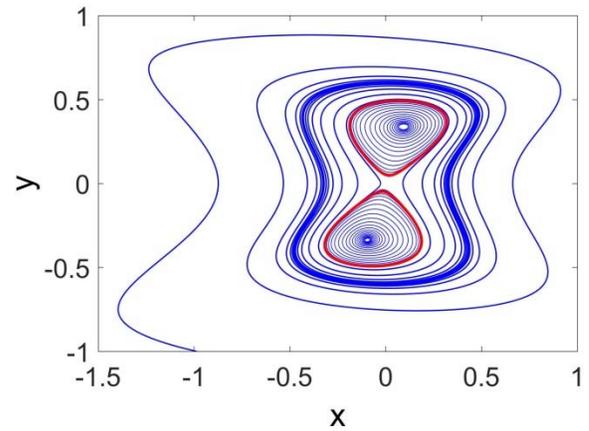
(b)

Figure 6. Comparison of the phase planes of systems (4) and (5) represented by fragments (a) and (b), respectively; $k = 0.31, \lambda = 0.1$.

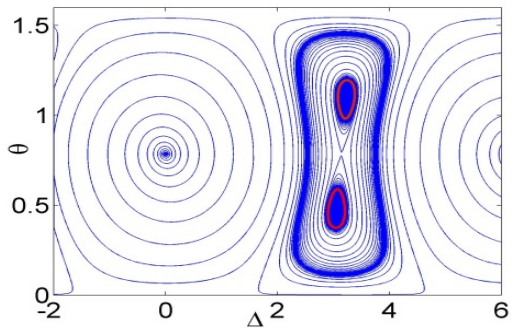
(a)

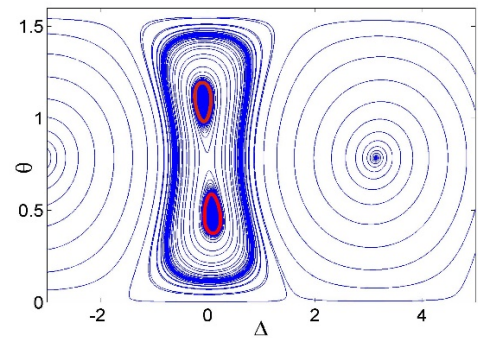
(b)

Figure 7. Comparison of the phase planes of system (4): a) 'stiff' nonlinearity case; (b) 'soft' nonlinearity case; $k = 0.31, \lambda = 0.1$.